\documentclass{svjour3}                 

\newcommand{\fd}{\emph{large-scale dataset}\xspace}
\newcommand{\sd}{\emph{reviewed dataset}\xspace}
\newcommand{\td}{\emph{developers dataset}\xspace}
\newcommand{\pd}{\emph{pre-training dataset}\xspace}

\usepackage{amsmath,amsfonts}
\usepackage{algorithm}
\usepackage{adjustbox}
\usepackage{lscape}
\usepackage{siunitx}
\usepackage[noend]{algpseudocode}
\usepackage{textcomp}
\usepackage{soul}
\usepackage{booktabs}
\usepackage[utf8]{inputenc}
\usepackage[T1]{fontenc}
\usepackage{rotating}
\usepackage{graphicx}
\usepackage{paralist}
\usepackage{tabularx}
\usepackage{multicol}
\usepackage{multirow}
\usepackage{pbox}
\usepackage{enumitem}	
\usepackage{colortbl}
\usepackage{pifont}
\usepackage{xspace}
\usepackage{url}
\usepackage{tikz}
\usepackage{tabularx}
\usepackage{fontawesome}
\usepackage{lscape}
\usepackage{listings}
\usepackage{color}
\usepackage{anyfontsize}
\usepackage{comment}
\usepackage{soul}
\usepackage{multibib}
\usepackage{multirow}
\usepackage{xspace}
\usepackage{caption}
\usepackage{footnote}
\usepackage{tcolorbox}
\usepackage{booktabs}
\usepackage{fixltx2e}
\usepackage{balance}
\usepackage[normalem]{ulem}
\usepackage{graphicx}

\definecolor{Gray}{gray}{0.9}
\definecolor{codegreen}{rgb}{0,0.6,0}
\definecolor{codegray}{rgb}{0.73,0.38,0.06}
\definecolor{codepurple}{rgb}{0.70,0.27,0}
\definecolor{codemagenta}{rgb}{0.74,0.09,0.42}
\definecolor{codeoutput}{rgb}{0.5,0,0}
\definecolor{backcolour}{rgb}{0.96,0.96,0.96}

\newboolean{showcomments}

\setboolean{showcomments}{true}

\ifthenelse{\boolean{showcomments}}
{\newcommand{\nb}[2]{
		\fbox{\bfseries\sffamily\scriptsize#1}
		{\sf\small$\blacktriangleright$\textit{#2}$\blacktriangleleft$}
	}
	
}
{\newcommand{\nb}[2]{}
	
}

\newcommand{\rev}[1]{\textcolor{black}{#1}}

\newcommand{\ie}{\emph{i.e.,}\xspace}
\newcommand{\eg}{\emph{e.g.,}\xspace}

\newcommand{\etal}{\emph{et~al.}\xspace}
\newcommand{\secref}[1]{Section~\ref{#1}\xspace}
\newcommand{\figref}[1]{Fig.~\ref{#1}\xspace}

\newcommand{\tabref}[1]{Table~\ref{#1}\xspace}

\begin{document}

\title{Automated Variable Renaming: Are We There Yet?}

\author{Antonio Mastropaolo         \and
        Emad Aghajani \and
        Luca Pascarella \and
        Gabriele Bavota.
}

\institute{Antonio Mastropaolo \at
              SEART @  Software Institute, Universit\`a della Svizzera italiana, Switzerland. \\
              \email{antonio.mastropaolo@usi.ch}         
           \and
          Emad Aghajani \at
          SEART @  Software Institute, Universit\`a della Svizzera italiana, Switzerland. \\
          \email{emad.aghajani@usi.ch}         
          \and
          Luca Pascarella \at
          SEART @  Software Institute, Universit\`a della Svizzera italiana, Switzerland. \\
          \email{luca.pascarella@usi.ch}         
          \and
          Gabriele Bavota \at
          SEART @  Software Institute, Universit\`a della Svizzera italiana, Switzerland. \\
          \email{gabriele.bavota@usi.ch}         
}

\date{Received: date / Accepted: date}

\maketitle

\begin{abstract}
Identifiers, such as method and variable names, form a large portion of source code. Therefore, low-quality identifiers can substantially hinder code comprehension. To support developers in using meaningful identifiers, several (semi-)automatic techniques have been proposed, mostly being data-driven (\eg statistical language models, deep learning models) or relying on static code analysis. Still, limited empirical investigations have been performed on the effectiveness of such techniques for recommending developers with meaningful identifiers, possibly resulting in rename refactoring operations. We present a large-scale study investigating the potential of data-driven approaches to support automated variable renaming. We experiment with three state-of-the-art techniques: a statistical language model and two DL-based models. The three approaches have been trained and tested on three datasets we built with the goal of evaluating their ability to recommend meaningful variable identifiers. Our quantitative and qualitative analyses show the potential of such techniques that, under specific conditions, can provide valuable recommendations and are ready to be integrated in rename refactoring tools. Nonetheless, our results also highlight limitations of the experimented approaches that call for further research in this field.


\end{abstract}

\section{Introduction} \label{sec:introduction}

Low-quality identifiers, such as meaningless method or variable names, are a recognized source of issues in software systems \cite{lin2019quality}. Indeed, choosing an expressive name for a program entity is not always trivial and requires both domain and contextual knowledge~\cite{carter1982choosing}. Even assuming a meaningful identifier is adopted in the first place while coding, software evolution may make the identifier not suitable anymore to represent a given entity. Moreover, the same entity used across different code components may be named differently, leading to inconsistent use of identifiers \cite{lin2017investigating}. For these reasons, rename refactoring has become part of developers' routine~\cite{murphy2011we}, as well as a standard built-in feature in modern integrated development environments (IDEs). While IDEs aid developers with the mechanical aspect of rename refactoring, developers remain responsible for identifying low-quality identifiers and choosing a proper rename.

To support developers in improving the quality of identifiers, several techniques have been proposed \cite{Thie2010a,Alla2014a,lin2017investigating,alon2019code2vec}. Among those, data-driven approaches are on the rise \cite{Alla2014a,lin2017investigating,alon2019code2vec}. This is also due to the recent successful application of these techniques in the code completion field \cite{Nguyen:icse2012,kim2020code,svyatkovskiy2020intellicode,Liu:ase2020,ciniselli2021empirical}, which is a more general formulation of the variable renaming problem. Indeed, if a model is able to predict the next code tokens that a developer is likely to write (\ie code completion), then it can be used to predict a token representing an identifier. Nevertheless, strong empirical evidence about the performance ensured by such data-driven techniques for supporting developers in identifier renaming is still minimal.

In this paper, we investigate the performance of three data-driven techniques in supporting automated variable renaming. We experiment with: (i) an $n$-gram cached language model \cite{hellendoorn2017deep}; (ii) the Text-to-Text Transfer Transformer (T5) model \cite{raffel2019exploring}, and (iii) the Transformer-based model presented by Liu \etal \cite{Liu:ase2020}. \rev{The $n$-gram cached language model \cite{hellendoorn2017deep} has been experimented against what were state-of-the-art deep learning models in 2017, showing its ability to achieve competitive performance in modeling source code. Thus, it represents a light-weight but competitive approach for the prediction of code tokens (including identifiers). Since then, novel deep learning models have been proposed such as, for example, those based on the transformer architecture \cite{vaswani2017attention}. Among those, T5 has been widely studied to support code-related tasks \cite{mastropaolo2021studying,mastropaolo2021icsme,tufano2022using,mastropaolo2022using,wang2021codet5,ciniselli2021tse}. Thus, it is representative of transformer-based models exploited in the literature. Finally, the approach proposed by Liu \etal \cite{Liu:ase2020} has been specifically tailored to improve code completion performance for identifiers, known to be among the most difficult tokens to predict. These three techniques provide a good representation of the current state-of-the-art of data-driven techniques for identifiers prediction.}

All experimented models require a training set to learn how to suggest ``meaningful'' identifiers. To this aim, we built a large-scale dataset composed of 1,221,193 instances, where an instance represents a \emph{Java} method with its related local variables. This dataset has been used to train, configure (\ie hyperparameters tuning), and perform a first assessment of the three techniques. In particular, as done in previous works related to the automation of code-related activities \cite{alon2019code2vec,Tufano:tosem2019,Tufano:icse2019,Watson:icse2020,haque:2020,tufano2021automating}, we considered a prediction generated by the models as correct if it resembles the choice made by the original developers (\ie if the recommended variable name is the same chosen by the developers). However, this validation assumes that the identifiers selected by the developers are meaningful, which is not always the case. 

To mitigate this issue and strengthen our evaluation, we built a second dataset using a novel methodology we propose to increase the confidence in the dataset quality (in our case, in the quality of the identifiers in code). In particular, we mined variable identifiers that have been introduced or modified during a code review process (\eg as a result of a reviewer's comment). These identifiers result from a shared agreement among multiple developers, thus increasing the confidence in their meaningfulness. This second dataset is composed of 457 \emph{Java} methods with their related local variables. 

Finally, we created a third dataset aimed at simulating the usage of the experimented tools for rename variable refactoring: We collected 400 Java projects in which developers performed a rename variable refactoring. By doing so, we were able to mine 442 valid commits. For each commit $c$ in our dataset, we checked-out the system's snapshot before ($s_{c-1}$) and after ($s_{c}$) the rename variable implemented in $c$. Given $v$ the variable renamed in $c$, we run the three techniques on the code in $s_{c-1}$ (\ie before the rename variable refactoring) to predict $v$'s identifier.

Then, we check whether the predicted identifier is the one implemented by the developers in $s_{c}$. If this is the case, this means that the approach was able to successfully recommend a rename variable refactoring for $v$, selecting the same identifier chosen by developers.

Our quantitative analysis shows that the Transformer-based model proposed by Liu \etal \cite{Liu:ase2020} is by far the best performing model in the literature for the task of predicting variable identifiers. This confirms the effort performed by the authors that aimed at specifically improve the performance of DL-based models in this task. This approach, named CugLM, can correctly predict the variable identifier in $\sim63\%$ of cases when tested on the large scale dataset we built. Concerning the other two datasets, the performance of all models drop, with CugLM still ensuring the best performance with $\sim45\%$ of correct predictions on both datasets.

We also investigate whether the ``confidence of the predictions'' generated by the three models (\ie how ``confident'' the model are about the generated prediction) can be used as a proxy for prediction quality. We found that when the confidence is particularly high ($>$ 90\%), the predictions generated by the models, and in particular by CugLM, have a very high chance of being correct ($>$80\% on the large-scale dataset). This suggests that the recommendations generated such tools, under specific conditions (\ie high confidence) are ready to be integrated in rename refactoring tools.  

We complement the study with a qualitative analysis aimed at inspecting ``wrong'' predictions to (i) see whether, despite being different from the original identifier chosen by the developer, they still represent meaningful identifiers for the variable provided as input; and (ii) distill lessons learned from these failure cases.

\rev{
	Concerning the first point, it is indeed important to clarify that even wrong predictions may be valuable for practitioners. This happens, for example, in the case in which the approach is able to recommend a valid alternative for an identifier (\eg \emph{surname} instead of \emph{lastName}) or maybe even suggesting a better identifier, thus implicitly recommending a rename refactoring operation. Such an analysis helps in better assessing the actual performance of the experimented techniques.
}

Finally, we analyze the circumstances under which the experimented tools tend to generate correct and wrong predictions. For example, not surprisingly, we found that these approaches are effective in recommending identifiers that they have already seen used, in a different context, in the training set. Also, the longer the identifier to predict (\eg in terms of number of terms composing it), the lower the likelihood of a correct prediction. 

\textbf{Significance of research contribution.} To the best of our knowledge, our work is the largest study at date experimenting with the capabilities of state-of-the-art data-driven techniques for variable renaming across several datasets, including two \emph{high-quality datasets} we built with the goal of increasing the confidence in the obtained results. The three datasets we built and the code implementing the three techniques we experiment with are publicly available \cite{replication}. Our findings unveil the potential of these tools as support for rename refactoring and help in identifying gaps that can be addressed through additional research in this field.

\section{Related Work} \label{sec:related}

We start by discussing works explicitly aiming at improving the quality of identifiers used in code. Then, we briefly present the literature related to code completion by discussing approaches that could be useful to suggest a variable name while writing code. For the sake of brevity, we do not discuss studies about the quality of code identifiers and their impact on comprehension activities (see \eg \cite{lawrie2006s,lawrie2007quantifying,butler2009relating,butler2010exploring}). 

\subsection{Improving the Quality of Code Identifiers}

Before diving into techniques aimed at improving the quality of identifiers, it is worth mentioning the line of research featuring approaches to split identifiers \cite{guerrouj2013tidier,corazza2012linsen} and expand abbreviations contained in them \cite{corazza2012linsen,hill2008amap,lawrie2011expanding}. These techniques can indeed be used to improve the expressiveness and comprehensibility of identifiers since, as shown by Lawrie \etal \cite{lawrie2006s}, developers tend to better understand expressive identifiers. 

On a similar research thread, Reiss \cite{reiss2007automatic} and Corbo \etal \cite{corbo2007smart} proposed tools to learn coding style from existing source code such that it can then be applied to the code under development. The learned style rules can include information about identifiers such as the token separator to use (\eg Camel or Snake case) and the presence of a prefix to name certain code entities (\eg \textit{OBJ\_varName}). The rename refactoring approaches proposed by Feldthaus and M{\o}ller \cite{feldthaus2013semi} and by Jablonski and Hou \cite{Jabl2007a}, instead, focus on the relations between variables, inferring whether one variable should be changed together with others.

Although the above-described methods can improve identifiers' quality (\eg by expanding abbreviated words and increasing consistency), they cannot address the use of meaningless/inappropriate identifiers as program entities' names.

To tackle this problem, Caprile and Tonella \cite{caprile2000restructuring} proposed an approach enhancing the meaningfulness of identifiers with a standard lexicon dictionary and a grammar collected by analyzing a set of programs, replacing non-standard terms in identifiers with a standard one from the dictionaries. 
 
Thies and Roth~\cite{thies2010recommending} and Allamanis \etal \cite{allamanis2014learning} also presented techniques to support the renaming of code identifiers. Thies and Roth~\cite{thies2010recommending} exploit static code analysis: if a variable \texttt{$v_1$} is assigned to an invocation of method \texttt{$m$} (\eg \texttt{name = getFullName}), and the type of \texttt{$v_1$} is identical to the type of the variable \texttt{$v_2$} returned by \texttt{$m$}, then rename \texttt{$v_1$} to \texttt{$v_2$}.

Allamanis \etal \cite{allamanis2014learning} proposed \textsc{NATURALIZE}, a two-step approach to suggest identifier names. In the first step, the tool extracts, from the code AST, a list of candidate names and, in the second step, it leverages an $n$-gram language model to rank the name list generated in the previous step. The authors evaluated the meaningfulness of the recommendations provided by their approach through analyzing 30 methods (for a total of 33 recommended variable renamings). Half of these suggestions were identified as relevant. Building on top of \textsc{NATURALIZE}, Lin \etal \cite{lin2017investigating} proposed \textsc{lear}, an approach combining code analysis and $n$-gram language models. The differences between \textsc{lear} and \textsc{NATURALIZE} are: (i) while \textsc{NATURALIZE} considers all the tokens in the source code, \textsc{lear} only focuses on tokens containing lexical information; (ii) \textsc{lear} also considers the type information of variables. Note that these techniques are meant to promote a consistent usage of identifiers within a given project (\ie renaming variables that represent the same \emph{entity} but are named differently within different parts of the same project). Thus, they cannot suggest naming a variable using an original identifier learned, for example, from other projects.

Differently from previous works, we empirically compare the effectiveness of data-driven techniques to support variable renaming, aiming to assess the extent to which developers could adopt them for rename refactoring recommendations. 

From this perspective, the most similar work to our study is the one by Lin \etal \cite{lin2017investigating}.
However, while their goal is to promote a consistent usage of identifiers within a project, we aim at supporting identifier rename refactoring in a broader sense, not strictly related to ``consistency''. Moreover, differently from Lin \etal \cite{lin2017investigating}, we include in our study recent DL-based techniques.

Finally, a number of previous works explicitly focused on the method renaming. Daka \etal \cite{Daka2017a} designed a technique to generate descriptive method names for automatically generated unit tests by summarizing API-level coverage goals. H{\o}st and {\O}stvold \cite{Host2009a} identify methods whose names do not reflect the responsibilities they implement. Alon \etal \cite{alon2019code2vec} proposed a neural model to represent code snippets and predict their semantic properties. Then, they use their approach to predict a method's name starting from its body. Differently from these works \cite{Host2009a,Daka2017a,alon2019code2vec}, we are interested in experimenting with data-driven techniques able to handle the renaming of variables.

\subsection{Automatic Code Completion}
While several techniques have been proposed to support code completion (see \eg \cite{Nguyen:icse2012,foster2012witchdoctor,ZhangYZFZZO12,Bruch:fse2009,ProkschLM15,niu2017api,svyatkovskiy2020fast,jin2018hidden,Robb2010a,svyatkovskiy2019pythia}), we only discuss approaches able to recommend/generate code tokens representing identifiers.

A precursor of modern code completion techniques is the \emph{Prospector} tool by Mandelin \etal \cite{MandelinXBK05}.  \emph{Prospector} aims at suggesting, within the IDE, variables or method calls from the user's codebase. Following this goal, other tools such as \emph{InSynth} by Gvero \etal \cite{GveroKKP13} and \emph{Sniff} by Chatterjee \etal \cite{chatterjee2009sniff} have added support for type completion (\eg expecting a type at a given point in the source code, the models search for type-compatible expressions).

Tu \etal \cite{Tu:fse2014} introduced a cache component that exploits code locality in $n$-gram models to improve their support to code completion. Results show that since code is locally repetitive, localized information can be used to improve performance. This enhanced model outperforms standard $n$-gram models by up to 45\% in terms of accuracy. On the same line, Hellendoorn and Devanbu \cite{Hellendoorn:fse2017} further exploit the cached models considering specific characteristics of code (\eg unlimited, nested, and scoped vocabulary). They also showed the superiority of their model when compared to deep learning for representing source code. This is one of the three techniques we experiment with (details in \secref{sub:ngram}). 

Karampatsis \etal \cite{Karampatsis:DLareBest}, a few years later, came to a different conclusion: Neural networks are the best language-agnostic algorithm for representing code. To overcome the \emph{out of vocabulary problem}, they propose the use of Byte Pair Encoding (BPE) \cite{bpe}, achieving better performance as compared to the cached $n$-gram model proposed in \cite{Hellendoorn:fse2017}. Also Kim \etal \cite{kim2020code} showed that novel DL techniques based on the Transformers neural network architecture can effectively support code completion. Similarly, Alon \etal \cite{alon2019structural} addressed the code completion problem with a language-agnostic approach leveraging the syntax to model the code snippet as a tree. Based on Long short-term Memory networks (LSTMs) and Transformers, the model receives an AST representing a partial statement with some missing tokens to complete. The versatility of addressing general problems with deep learning models pushed researchers to explore the potential of automating various code-related tasks. 

For example, Svyatkovskiy \etal \cite{svyatkovskiy2020intellicode} created \emph{IntelliCode Compose}, a general-purpose code completion tool capable of predicting code sequences of arbitrary tokens. Liu \etal \cite{Liu:ase2020} presented a Transformer-based neural architecture pre-trained on two development tasks: code understanding and code generation. Successively, the model is specialized with a fine-tuning step on the code completion task. With this double training process, the authors try to incorporate a generic knowledge of the project into the model with the explicit goal of improving performance in predicting code identifiers. This is one of the techniques considered in our study (details in \secref{sub:cuglm_ase}).

Ciniselli \etal~\cite{ciniselli2021empirical} demonstrated how deep learning models can support code completion at different granularities, not only predicting inline tokens but even complete statements. To this aim, they adapt and train a BERT model in different scenarios ranging from simple inline predictions to entire blocks of code.

Finally, Mastropaolo \etal~\cite{mastropaolo2021studying}\footnote{In the following we refer to \cite{mastropaolo2021studying} as Mastropaolo \etal because the set of authors only partially overlaps.} recently showed that T5~\cite{raffel2019exploring}, properly pre-trained and fine-tuned, achieves state-of-the-art performance in many code-related tasks, including bug-fixing, mutants injection, code comment and assert statement generation. This is the third approach considered in our study (details in \secref{sub:t5}). 

\rev{Companies have also shown active interest in supporting practitioners while developing sources with code completion tools. For example, IntelliCode was introduced by Svyatkovskiy \etal~\cite{svyatkovskiy2020intellicode}, a group of researchers working at the Microsoft Corporation. It is multilingual code completion tool that aims at predicting sequences of code tokens such as entire lines of syntactically correct code. IntelliCode leverages a transformer model and has been trained on 1.2 billion lines
of diverse programming languages. Similarly, GitHub Copilot has been recently introduced as a state-of-the-art code recommender~\cite{copilot,howard2021github}. It has been experimentally released through their API that aims at translating natural language descriptions into source code~\cite{chen2021evaluating}. Copilot is based on a GPT-3 model \cite{GPT3} fine-tuned on publicly available code from GitHub.} Nonetheless, the exact dataset used for its training is not publicly available. This hinders the possibility to use it in our study, since we cannot check for possible overlaps between the Copilot’s training set and the test sets used in our study (also collected from GitHub open source projects).

The successful applications of DL for code completion suggest its suitability for supporting rename refactoring. Indeed, if a model is able to predict the identifier to use in a given context, it can be used to boost identifiers' quality as well. As a representative of DL-based techniques to experiment with variable rename refactoring task, we chose (i) the model by Liu \etal~\cite{Liu:ase2020}, given its focus on the predicting code identifiers; and (ii) the T5 model as used in~\cite{mastropaolo2021studying}, given its state-of-the-art performance across many code-related tasks. We acknowledge that other choices are possible given the vast literature in the field. We focused on two DL-models recently published in top software engineering venues (\ie ASE'20~\cite{Liu:ase2020} and ICSE'21~\cite{mastropaolo2021studying}).


\section{Data-driven Variable Renaming} \label{sec:techniques}
In our study, we aim at assessing the effectiveness of data-driven techniques for automated variable renaming.
We focus on three techniques representative of the state-of-the-art. The first is a statistical language model that showed its effectiveness in modeling source code \cite{hellendoorn2017deep}. The second, T5 \cite{raffel2019exploring}, is a recently proposed DL-based technique already applied to address code-related tasks~\cite{mastropaolo2021studying}. The third is the Transformer-based model presented by Liu \etal~\cite{Liu:ase2020}  to boost code completion performance on identifiers. 

\begin{figure}[!ht]
\centering
	\includegraphics[width=\linewidth]{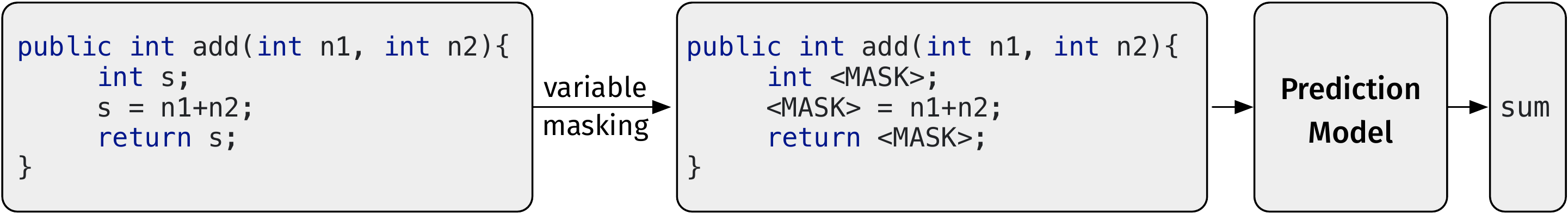}
	\caption{Variable renaming scenario}
	\label{fig:task}
\end{figure}

\figref{fig:task} depicts the scenario in which these techniques have been experimented. We work at method-level granularity: For each local variable $v$ declared in a method $m$, we mask every $v$'s reference in $m$ asking the experimented techniques to recommend a suitable name for $v$. If the recommended name is different from the original one, a rename variable recommendation can be triggered.

We provide an overview of the experimented techniques, pointing the reader to the papers introducing them \cite{hellendoorn2017deep,raffel2019exploring,Liu:ase2020} for additional details. Our implementations are based on the ones made available by the original authors of these techniques and are publicly available in our replication package \cite{replication}. The training of the techniques is detailed in \secref{sec:design}.

\subsection{N-gram Cached Model} \label{sub:ngram}
Statistical language models can assess a probability of a given sequence of words. The basic idea behind these models is that the higher the probability, the higher the ``\emph{familiarity}'' of the scored sequence. Such \emph{familiarity} is learned by training the model on a large text corpus. An $n$-gram language model predicts a single word following the $n-1$ words preceding it. In other words, $n$-gram models assume that the probability of a word only depends on the previous $n-1$ words. 

Hellendoorn and Devanbu \cite{hellendoorn2017deep} discuss the limitations of $n$-gram models that make them suboptimal for modeling code (\eg the unlimited vocabulary problem due to new words that developers can define in identifiers). To overcome these limitations, the authors present a \emph{dynamic, hierarchically scoped, open vocabulary language model} \cite{hellendoorn2017deep}, showing that it can outperform Recurrent Neural Networks (RNN) and LSTM in modeling code. While Karampatsis \etal \cite{Karampatsis:DLareBest} showed that DL models can outperform the cached $n$-gram model, the latter ensures good performance at a fraction of the DL models training cost, making it a competitive baseline for code-related tasks. 

\subsection{Text-To-Text-Transfer-Transformer (T5)} \label{sub:t5}
The T5 model has been introduced by Raffel \etal \cite{raffel2019exploring} to support multitask learning in Natural Language Processing. 
The idea is to reframe NLP tasks in a unified text-to-text format in which the input and output of all tasks to support are always text strings. For example, a single T5 model can be trained to translate across a set of different languages (\eg English, German) and identify the sentiment expressed in sentences written in any of those languages. This is possible since both these tasks (\ie translation and sentiment identification) are text-to-text tasks, in which a text is provided as input (\ie a sentence in a specific language for both tasks) and another text is generated as output (\ie the translated sentence or a label expressing the sentiment). T5 is trained in two phases: \textit{pre-training}, which allows defining a shared knowledge-base useful for a large class of text-to-text tasks (\eg guessing masked words in English sentences to learn about the language), and \textit{fine-tuning}, which specializes the model on a specific downstream task (\eg learning the translation of sentences from English to German).
As previously said, T5 already showed its effectiveness in code-related tasks \cite{mastropaolo2021studying}. However, its application to variable renaming is a premier. Among the T5 variants proposed by Raffel \etal~\cite{raffel2019exploring} that mostly differ in terms of architectural complexity, we adopt the smallest one (T5\textsubscript{\textit{small}}).
The choice of such architecture is driven by our limited computational resources. However, we acknowledge that bigger models have been shown to further increase performance~\cite{raffel2019exploring}.

\subsection{Deep-Multi-Task code completion model}\label{sub:cuglm_ase}

Liu \etal \cite{Liu:ase2020} recently proposed the \emph{Code Understanding and Generation pre-trained Language Model (CugLM)}, a BERT-based model for source code modeling. Albeit under-the-hood CugLM still features a Transformer-based network \cite{attention} as T5, such an approach has been specifically conceived to improve the performance of language models in identifiers, thus making it very suitable for our study on variable renaming. CugLM is pre-trained using three objectives. The first asks the model to predict masked identifiers in code (being thus similar to the one used in the T5 model, but focused on identifiers). The second task asks the model to predict whether two fragments of code can follow each other in a snippet. Finally, the third is a left-to-right language modeling task, in which the classic code completion scenario is simulated, \ie given some tokens (left part), guess the following token (right part). 

Once pre-trained, the model is fine-tuned for code completion in a multi-task learning setting, in which the model has first to predict the type of the following token and, then, the predicted type is used to foster the prediction of the token itself. As reported by the authors, such an approach achieves state-of-the-art performance when it comes to predicting code identifiers.
\section{Study Design} \label{sec:design}

The \emph{goal} is to experiment the effectiveness of data-driven techniques in supporting automated variable renaming. The \emph{context} is represented by (i) the three techniques \cite{hellendoorn2017deep,raffel2019exploring,Liu:ase2020} introduced in \secref{sec:techniques} and (ii) three datasets we built for training and evaluating the approaches. Our study answers the following research question: \textbf{\emph{To what extent can data-driven techniques support automated variable renaming?}}

\subsection{Datasets Creation}
\label{sub:datasets}

To train and evaluate the experimented models, we built three datasets: (i) the \fd, used to train the models, tune their parameters, and perform a first assessment of their performance; (ii) the \sd and (iii) the  \td used to further assess the performance of the experimented techniques. Our quantitative evaluation is based on the following idea: If, given a variable, a model is able to recommend the same identifier name as chosen by the original developers, then the model has the potential to generate meaningful rename recommendations. 

Clearly, there is a strong assumption here, namely that the identifier selected by the developers is meaningful. For this reason, we have three datasets. The first one (\fd) aims at collecting a high number of variable identifiers that are needed to train the data-driven models and test them on a large number of data points. The second one (\sd) focuses instead on creating a test set of high-quality identifiers for which our assumption can be more safely accepted: These are identifiers that have been modified or introduced during a code review process. Thus, more than one developer agreed on the appropriateness of the chosen identifier name for the related variable. Finally, the third dataset (\td) focuses on identifiers that have been subject to a rename refactoring operation (\ie the developer put effort in improving the quality of the identifier through a refactoring). Again, this increases our confidence in the quality of the considered identifiers.

In this section, we describe the datasets we built, while \secref{sub:training} details how they have been used to train, tune, and evaluate the three models.

\subsubsection{Large-scale Dataset}
\label{sub:large-scale}
We selected projects on GitHub \cite{github} by using the search tool by Dabic \etal \cite{dabic2021sampling}. This tool indexes all GitHub repositories written in 13 different languages and having at least 10 stars, providing a handy querying interface \cite{ghs} to identify projects meeting specific selection criteria. We extracted all \emph{Java} projects having at least 500 commits and at least 10 contributors. We do so as an attempt to discard toy/personal projects. We decided to focus on a single programming language to simplify the toolchain building needed for our study. Also, we excluded forks to reduce the risk of duplicated repositories in our dataset. 

Such a process resulted in 5,369 cloned Java projects from which we selected the 1,425 using Maven\footnote{Maven is a software project management tool that, as reported in its official webpage (\url{https://maven.apache.org}) ``can manage a project's build, reporting and documentation from a central piece of information''.}  \cite{maven} and having their latest snapshot being compilable.  \rev{Maven allows to quickly verify the compilability of the projects, which is needed to extract information about types needed by one of the experimented models (\ie \emph{CugLM}). \emph{CugLM} leverages identifiers' type information to improve its predictions. To be precise and comprehensive in type resolution, we decided to rely on the JavaParser library \cite{javaparser}, running it on compilable projects. This allows to resolve also types that are implemented in imported libraries. We provide the tool we built for such an operation as part of our replication package~\cite{replication}.}

We used \emph{srcML} \cite{SrcML} to extract from each \emph{Java} file contained in the 1,425 projects all methods having \textit{\#tokens $\leq$ 512}, where \textit{\#tokens} represents the number of tokens composing a function (excluding comments). The filter on the maximum length of the method is needed to limit the computational expense of training DL-based models (similar choices have been made in previous works \cite{Tufano:tosem2019,haque:2020,tufano2021automating}, with values ranging between 50 and 100 tokens). All duplicate methods have been removed from the dataset to avoid overlap between training and test sets we built from them.

From these 1,425 repositories, we set apart 400 \rev{randomly selected} projects for constructing the \td (described in \secref{sub:dev-dataset}). Concerning the remaining 1,025, we use $\sim$40\% of them (418 randomly picked repositories) to build a dataset needed for the pre-training of the T5 \cite{raffel2019exploring} and of the \emph{CugLM} model \cite{Liu:ase2020} (\pd). 
\rev{Such a dataset is needed to support the pre-training phase that, as shown in the literature, helps deep learning models to achieve better performance when dealing with code-related tasks \cite{tufano2020unit,tufano2022using,mastropaolo2022using,ciniselli2021empirical}. Indeed, the pre-training phase conveys two major advantages summarized as follows: (i) once the model has been pre-trained, it can learn general representations and patterns of the language the model is working with, (ii) the pre-trained model yields to a more robust model initialization of the neural network weights that can then support the specialization phase (\ie fine-tuning). 
}

The remaining 615 projects (\fd) have been further split into training (60\%), evaluation (20\%), and test (20\%). The training set has been used to fine-tune the two DL-based models (\ie T5 and \emph{CugLM}). This dataset, joined with the \pd, has also been used to train the $n$-gram model. In this way, all models have been trained using the same set of data, with the only difference being that the training is organized in two steps (\ie pre-training and fine-tuning) for T5 and \emph{CugLM}, while it consists of a single step for the $n$-gram model.

For the T5 model, we used the evaluation set to tune its hyperparameters (\secref{sub:training}), since no previous work applied such a model for the task of variable renaming. Instead, for \emph{CugLM} and $n$-gram we used the best configurations reported in the original works presenting them \cite{hellendoorn2017deep,Liu:ase2020}. Finally, the test set has been used to perform a first assessment of the models' performance.

\begin{table}[h]
	\centering
	\caption{Num. of methods in the datasets used in our study}
	\begin{tabular}{lrrr}
		\toprule
		\textbf{Dataset} & \textbf{train}    & \textbf{eval}     & \textbf{test} \\
		\midrule
		\pd & 500,414 & - & -\\
		\fd                           & 394,574            & 176,944    &149,261            \\
		\sd                           & -             					 & -  				 & 		457\\
		\td                           & -             					 & -  				 & 			442              \\
		\bottomrule
	\end{tabular}
	\label{tab:datasets}
	
\end{table}

\tabref{tab:datasets} shows the size of the datasets in terms of the number of extracted methods (\sd and \td are described in the following).

\subsubsection{Reviewed Dataset}
Also in this case, we selected GitHub projects using the tool by Dabic \etal \cite{dabic2021sampling}. Since the goal for the \sd is to mine code review data, we added on top of the selection criteria used for the \fd a minimum of 100 pull requests per selected project. Also in this case we only selected Maven projects having their latest snapshot successfully compiling. We then mined from the 948 projects we obtained information related to the code review performed in their pull requests. Let us assume that a set of files \textit{C\textsubscript{s}} is submitted for review. A set of reviewer comments ${R\textsubscript{c}}$ can be made on \textit{C\textsubscript{s}} possibly resulting in a revised version of the code \textit{C\textsubscript{r$_1$}}. 

Such a process is iterative and can consists of several rounds each one generating a new revised version \textit{C\textsubscript{r$_i$}}. Eventually, if the code contribution is accepted for merging, this concludes the review process with a set of \textit{C\textsubscript{f}} files. This whole process ``transforms'' \textit{C\textsubscript{s}} $\rightarrow$ \textit{C\textsubscript{f}}. We use \emph{srcML} to extract from both \textit{C\textsubscript{s}} and \textit{C\textsubscript{f}} the list of methods in them and, by performing a \emph{diff}, we identify all variables that have been introduced or modified in each method as result of the review process (\ie all variables that were not present in \textit{C\textsubscript{s}} but that are present in \textit{C\textsubscript{f}}). We conjecture that the identifiers used to name these variables, being the output of a code review process, have a higher chance of representing high quality data that can be used to assess the performance of the experimented models.

Also in this case we removed duplicate methods both (i) within the \sd, and (ii)  between it and the previous ones (\pd and training set of \fd ), obtaining 457 methods usable as a further test set of the three techniques.

\subsubsection{Developers' Dataset} \label{sub:dev-dataset}
We run \emph{Refactoring miner} \cite{refminer:tse} on the history of the 400 Java repositories we previously put aside. Refactoring miner is the state-of-the-art tool for refactoring detection in Java systems. We run it on every commit performed in the 400 projects, looking for commits in which a \emph{Rename Variable refactoring} has been performed on the local variable of a method. 
This gives us, for a given commit $c_i$, the variable name at commit  $c_{i-1}$ (\ie before the refactoring) and the renamed variable in commit $c_i$. We use this set of commits as an additional test set (\td) to verify if, by applying the experimented techniques on the $c_{i-1}$ version, they are potentially able to recommend a rename (\ie they suggest, for the renamed variable, the identifier applied with the rename variable refactoring). After removing duplicated methods from this dataset as well (similarly, we also removed duplicates between \td, \pd, and the training set in \fd), we ended up with 442 valid instances.

\subsection{Training and Hyperparameters Tuning of the Techniques}
\label{sub:training}

\subsubsection{\textbf{N-gram model}} The $n$-gram model has been trained on the instances in \pd  $\cup$ \fd. This means that all Java methods contained in \pd and in the training set of \fd have been used for learning the probability of sequences of tokens. We use $n=3$ since higher values of $n$ have been proven to result in marginal performance gains \cite{hellendoorn2017deep}.  

\subsubsection{\textbf{T5}} To pre-train the T5, we use a self-supervised task similar to the one by Raffel \etal \cite{raffel2019exploring}, in which we randomly mask $15\%$ of code tokens in each instance (\ie a Java method) from the \pd, asking the model to guess the masked tokens. Such a training is intended to give the model general knowledge about the language, such that it can perform better on a given \emph{down-stream} task (in our case, guessing the identifier of a variable). The pre-training has been performed for 200k steps (corresponding to $\sim$13 epochs on our \pd) since we did not observe any improvement going further. We used a 2x2 TPU topology (8 cores) from Google Colab to train the model with a batch size of 128. As a learning rate, we use the \emph{Inverse Square Root} with the canonical configuration \cite{raffel2019exploring}. We also created a new \emph{SentencePiece} model (\ie a tokenizer for neural text processing) by training it on the entire pre-training dataset so that the T5 model can properly handle the Java language. We set its size to 32k word pieces.

In order to find the best configuration of hyper-parameters, we rely on the same approach used by Mastropaolo \etal \cite{mastropaolo2021studying}. Specifically, we do not tune the hyperparameters of the T5 model for the pre-training (\ie we use the default ones), because the pre-training itself is task-agnostic, and tuning may provide limited benefits. Instead, we experiment with four different learning rate schedulers for the fine-tuning phase. Since this is the first time T5 is used for recommending identifiers, we also perform an ablation study aimed at assessing the impact of pre-training on this task. Thus, we perform the hyperparameter tuning for both the pre-trained and the non pre-trained model, experimenting with the four configurations in \tabref{tab:hyperT5}: constant (C-LR), slanted triangular (ST-LR), inverse square (ISQ-LR), and polynomial (PD-LR) learning rate. We experiment the same configurations for the pre-trained and the non-pretrained models, with the only difference being the $\mathit{LR}_{\mathit{starting}}$ and $\mathit{LR}_{\mathit{end}}$ of the PD-LR. Indeed, in the non pre-trained model (ablation study), we had to lower those values to make the gradient stable (see \tabref{tab:hyperT5}).

\begin{table}[h]
	\centering
	\small
	\caption{Hyperparameters tuning for the T5 Model}
	\begin{tabular}{llrr}
		\toprule
		\textbf{Learning Rate}   & \textbf{Parameters} & \textbf{Pre-Trained} & \textbf{No Pre-Trained} \\
		\midrule
		 C-LR            & $\mathit{LR}$       & $0.001$ & $0.001$  \\
		 ST-LR  
		& $\mathit{LR}_{\mathit{starting}}$ & $ 0.001$ & $ 0.001$ \\
		& $\mathit{LR_{\mathit{max}}}$ & $ 0.01$ & $ 0.01$ \\
		& $\mathit{Ratio}$ & $32$ & $32$ \\
		& $\mathit{Cut}$ & $0.1$ & $0.1$\\
		 ISQ-LR 
		& $\mathit{LR}_{\mathit{starting}}$ & $0.01$ & $0.01$ \\
		& $\mathit{Warmup}$ & $10,000$ & $10,000$ \\
		 PD-LR    
		& $\mathit{LR}_{\mathit{starting}}$ & $0.01$ & $0.001$ \\
		& $\mathit{LR}_{\mathit{end}}$ & $0.01$ & $0.001$ \\
		& $\mathit{Power}$ & $0.5$ & $0.5$\\
		\bottomrule
	\end{tabular}
	
	\label{tab:hyperT5}
\end{table}

We fine-tune the T5 for 100k steps for each configuration. Then, we compute the percentage of correct predictions (\ie cases in which the model can correctly predict the masked variable identifier) achieved in the evaluation set. The achieved results reported in \tabref{tab:hyperparameter_results} showed a superiority of the ST-LR (second column) for the non pre-trained model, while for the pre-trained model, the PD-LR works slightly better. Thus, we use these two scheduler in our study for fine-tuning the final models for 300k steps.

\begin{table}[h!]
	\centering
	\caption{T5 hyperparameter tuning results}
	\begin{tabular}{lrrrr}
		\toprule
		\textbf{Experiment}                  & \textbf{C-LR}    & \textbf{ST-LR}      & \textbf{ISQ-LR}        & \textbf{PD-LR} \\
		\midrule
		Pre-trained                            &  30.74\%                & 29.11\%    		  & 30.77\%                  &  \textbf{30.80}\%         \\
		No Pre-trained                        		&  21.18\%                & \textbf{27.56\%}      	   &  26.08\%                  &  23.90\%         \\		
		\bottomrule
	\end{tabular}
	\label{tab:hyperparameter_results}
\end{table}

The fine-tuning of the T5 required some further processing to the \fd. Given a Java method $m$ having $n$ distinct local variables, we create $n$ versions of it $m_1$, $m_2$, $\dots$, $m_n$ each one having all occurrences of a specific variable masked with a special token. Such a representation of the dataset allows to fine-tune the T5 model by providing it pairs ($m_j$, $i_j$), where $m_j$ is a version of $m$ having all occurrences of variable $v_j$ replaced with a \texttt{<MASK>} token and $i_j$ is the identifier selected by the developers for $v_j$. This allows the T5 to learn proper identifiers to name variables in specific code contexts. The same approach has been applied on the \fd evaluation and test set, as well as on the \sd and \td. In these cases, an instance is a method with a specific variable masked, and the trained model is used to guess the masked identifier. \tabref{tab:t5_datasets} reports the number of instances in the datasets used for the T5 model. Note that such a masking processing was not needed for the $n$-gram model nor for \emph{CugLM}, since they just scan the code tokens during training, and they try to predict each code token sequentially during testing. Still, it is important to highlight that all techniques have been trained and tested on the same code. 

\begin{table}[ht!]
	\centering
	\caption{Instances in the datasets used for training, evaluating, testing the T5 model\vspace{-0.2cm}}
	\begin{tabular}{lrrr}
		\toprule
		                  				& \textbf{train}    & \textbf{eval}     & \textbf{test} \\
		\midrule
		\fd                           &  1,122,864           	& 521,779  				& 	437,384            \\
		\sd                           & -             					 & -  				 & 457 \\
		\td                           & -             					 & -  				 & 442 \\
		\bottomrule
	\end{tabular}
	\label{tab:t5_datasets}
\end{table}

\subsubsection{\textbf{CugLM}} To pre-train and fine-tune the \emph{CugLM} model we first retrieved the identifiers' type information for all code in the \pd and \fd. Then, we leveraged the script provided by the original authors in the replication package \cite{cuglm} to obtain the final instances in the format expected by the model. For both pre-training and fine-tuning (described in \secref{sub:cuglm_ase}), we rely on the same hyper-parameters configuration used by the authors in the paper presenting this technique \cite{Liu:ase2020}.

\subsection{Performance Assessment} \label{sub:test}
We assess the performance of the trained models against the \textit{large-scale test set}, the \sd, and the \td. 
For each prediction made by each model, we collect a measure acting as ``confidence of the prediction'', \ie a real number between 0.0 and 1.0 indicating how confident the model is about the prediction. For the $n$-gram model, such a measure is a transformation of the entropy of the predictions. Concerning the T5, we exploited the \texttt{score} function to assess the model's confidence on the provided input. The value returned by this function ranges from minus infinity to 0 and it is the log-likelihood ($ln$) of the prediction. Thus, if it is 0, it means that the likelihood of the prediction is 1 (\ie the maximum confidence, since $ln(1) = 0$), while when it goes towards minus infinity, the confidence tends to 0. Finally, CugLM outputs the \emph{log-prob} for each predicted tokens. Hence, we normalize this value throught the \emph{exp} function.

We investigate whether the confidence of the predictions represents a good proxy for their quality. If the confidence level is a reliable indicator of the predictions' quality (\eg 90\% of the predictions having $c > 0.9$ are correct), it can be extremely useful in the building of recommender systems aimed at suggesting rename refactorings, since only recommendations with high confidence could be proposed to the developer.
We split the predictions by each model into ten intervals, based on their confidence $c$ going from 0.0 to 1.0 at steps of 0.1 (\ie first interval includes all predictions having 0 $\leq$ c $<$ 0.1, last interval has 0.9 $\leq$ c). Then, we report for each interval the percentage of \emph{correct predictions} generated by each model in each interval. To assess the performance of the techniques overall, we also report the percentage of correct predictions generated by the models on the entire test datasets (\ie by considering predictions at any confidence level).

\begin{figure}[h!]
\centering
	\includegraphics[width=0.6\linewidth]{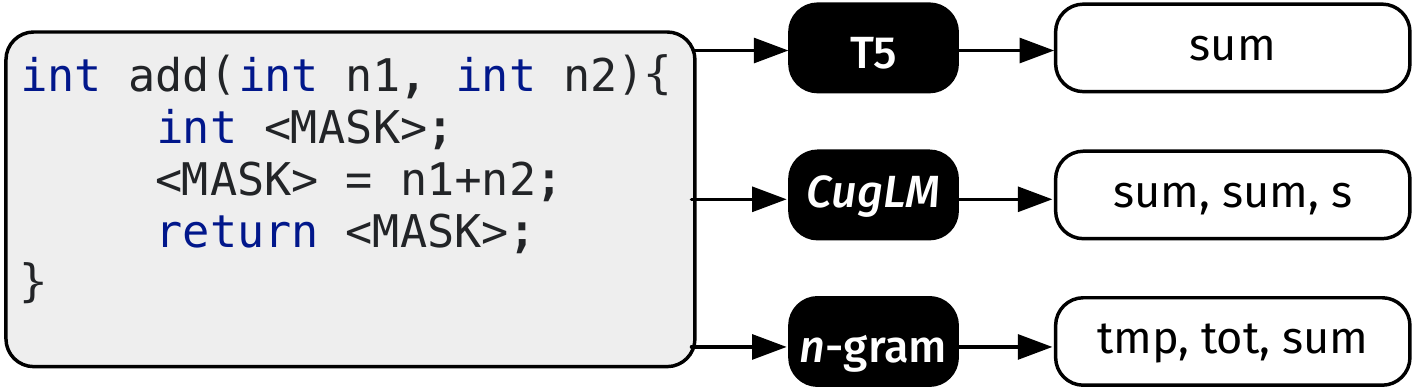}
	\caption{Example of models' outputs for a test instance}
	\label{fig:output}
\end{figure}

A prediction is considered ``correct'' if the predicted identifier corresponds to the one chosen by developers in the \fd and in the \sd, and if it matches the renamed identifier in the \td. However, a clarification is needed on the way we compute the correct predictions. We explain this process through \figref{fig:output}, showing the output of the experimented models, given an instance in the test sets.

The grey box in \figref{fig:output} represents an example of an instance in the test set: a function having all references to a specific variable originally named \texttt{sum} masked. The T5 model, given such an instance as input, predicts a \textbf{single} identifier (\ie \texttt{sum}) for all three references of the variable. Thus, for the T5, it is easy to say whether the single generated prediction is equivalent to the identifier chosen by the developers or not. The $n$-gram and the \emph{CugLM} model, instead, generate three predictions, one for each of the masked instances, despite  they represent the same identifier. 

Thus, for these two models, we use two approaches to compute the percentage of correct predictions in the test sets. The first scenario, named \emph{complete-match}, considers the prediction as correct only if all three references to the variable are correctly predicted. Therefore, in the example in \figref{fig:output}, the prediction of the \emph{CugLM} model (2 out of 3) is considered wrong. Similarly, the $n$-gram prediction (1 out of 3 correct) is considered wrong. The second scenario, named \emph{partial-match}, considers a prediction as correct if at least one of the instances is correctly predicted (thus, in \figref{fig:output} both the $n$-gram and the \emph{CugLM} predictions are considered correct). 

We also statistically compare the performance of the models in terms of correct predictions: We use the McNemar's test \cite{mcnemar}, which is a proportion test suitable to pairwise compare dichotomous results of two different treatments. We statistically compare each pair of techniques in our study (\ie T5 \emph{vs} \emph{CugLM}, T5 \emph{vs} $n$-gram, \emph{CugLM} \emph{vs} $n$-gram). To compute the test results for two techniques $T_1$ and $T_2$, we create a confusion matrix counting the number of cases in which (i) both $T_1$ and $T_2$ provide a correct prediction, (ii) only $T_1$ provides a correct prediction, (iii) only $T_2$ provides a correct prediction, and (iv) neither $T_1$ nor $T_2$ provide a correct prediction. We complement the McNemar's test with the Odds Ratio (OR) effect size. Also, since we performed multiple comparisons, we adjusted the obtained $p$-values using the Holm's correction \cite{Holm1979a}.

We also manually analyzed a sample of wrong predictions generated by the approaches with the goal of (i) assessing whether, despite being different from the original identifiers used by the developers, they were still meaningful; and (ii) identifying scenarios in which the experimented techniques fail.  \rev{To perform such an analysis, we selected the top-100 wrong predictions for each approach (300 in total) in terms of confidence level. Three of the authors inspected all of them, trying to understand if the generated variable name could have been a valid alternative for the target one, while the fourth author solved conflicts. Note that, given 100 wrong predictions inspected for a given model, we do not check whether the other models correctly predict these cases. This is not relevant for our analysis, since the only goal is to understand the extent to which the ``wrong'' predictions generated by each model might still be valuable for developers.}

Finally, we compare the correct and wrong predictions in terms of (i) the size of the context (\ie number of tokens composing the method and number of times a variable is used in the method), (ii) the length of the identifier in terms of number of characters and number of words composing it (as obtained through camelCase splitting); (iii) the number of times the same identifier appears in the training data. We show these distributions using boxplots comparing the two groups of predictions (\eg compare the length of identifiers in correct and wrong predictions). We also compare these distributions using the Mann-Whitney test \cite{Conover:1998} and the Cliff's Delta ($d$) to estimate the magnitude of the differences \cite{Cliff:2005}. We follow well-established guidelines to interpret the effect size: negligible for $|d| < 0.10$, small for $0.10 \le |d| < 0.33$, medium for $0.33 \le |d| < 0.474$, and large for $|d| \ge 0.474$ \cite{Cliff:2005}.
\section{Results Discussion} \label{sec:results}

\tabref{tab:correct} reports the results achieved by the three experimented models for each dataset in terms of correct predictions. For T5, both pre-trained and non pre-trained versions are presented. For the $n$-gram and CugLM, we report the results both when using \textit{perfect match} and the \textit{partial match} heuristic to compute the correct predictions, while this was not required for the T5 for which the results should be interpreted as \textit{perfect matches} (see \secref{sub:test}).

Before commenting on the results is also important to clarify that the cached $n$-gram model \cite{hellendoorn2017deep} exploited, as compared to the other two models, additional information due to the caching mechanism. Indeed, the caching allows the model to ``look'' at code surrounding the one for which tokens must be predicted (in our case, the method in which we want to predict the variable identifier). Given a method $m_t$ in the test set, we provide its cloned repository as ``test folder'' to the $n$-gram model, in such a way that it is leveraged by the caching mechanism (we used the implementation from \cite{hellendoorn2017deep}).

Two observations can be easily made by looking at \tabref{tab:correct}. First, for T5, the pre-trained model works (as expected) better than its non pre-trained version. From now on, we focus on the pre-trained T5 in the discussion of the results. Second, consistently for all three datasets, CugLM outperforms the other models by a significant margin. In particular, when looking at the correct predictions (\emph{complete match}), the improvement is +26\% and +53\% over T5 and $n$-gram, respectively, in the \fd. The gap is smaller but still substantial for the \sd (+15\% and +44\% for T5 and $n$-gram) and for the \td (+22\% and +43\%). The difference in performance in favor of CugLM is always statistically significant (see \tabref{tab:stats-correct}), with ORs going from 1.98 to 98.0. For example, on the \fd the ORs indicate that CugLM has 3.54 and 23.06 higher odds to generate a correct prediction as compared to the T5 and the $n$-gram model. These results confirm the suitability of the model proposed by Liu \etal \cite{Liu:ase2020} when it comes to predicting code identifiers. 

\tabref{tab:correct} also shows that, as expected by construction, the percentage of correct predictions generated by CugLM and by the $n$-gram model increases when considering the \textit{partial match} heuristic. However, for a fair comparison with the T5 model, we mostly focus our discussion on the \textit{perfect match} scenario, that is also the one used in the computation of the statistical tests (\tabref{tab:stats-correct}).

The trend in performance is the same across the three datasets. However, the accuracy of all models drops on the \sd and on the \td. Still, even in this scenario, CugLM is able to correctly recommend $\sim$50\% of identifiers.

\begin{landscape}
\begin{table*}
	\centering
	\caption{Correct predictions: C-match indicates the \emph{complete-match} heuristic, P-match the \emph{partial-match}\vspace{0.2cm}}
	\label{tab:correct}
	\setlength\extrarowheight{13pt}
	\begin{adjustbox}{width=\columnwidth}
		\begin{tabular}{crrrrrrrrrrrrrrrrr}
			\toprule
			\multicolumn{17}{c}{{\bf \fd}}\\\midrule
			\multirow{2}{*}{\#Instances}  & \multicolumn{2}{c}{{n-gram} (c-match)} && \multicolumn{2}{c}{ {n-gram (p-match) }} 
			&& \multicolumn{2}{c}{ {CugLM (c-match) }} && \multicolumn{2}{c}{ {CugLM (p-match) }} 
			&&\multicolumn{2}{c}{ {T5 (non pre-trained) }} && \multicolumn{2}{c}{ {T5 (pre-trained) }} 
			\\ \cline{2-3} \cline{5-6} \cline{8-9} \cline{11-12} \cline{14-15} \cline{17-18}
			& {\#Correct} & {\%Correct} &&  {\#Correct} & {\%Correct}  && {\#Correct} & {\%Correct} && {\#Correct} & {\%Correct}  && {\#Correct} & {\%Correct} && {\#Correct} & {\%Correct}  \\ \hline
			437,384  & 46,126 & 10.54\% && 167,868 & 38.38\%&& 277,595 & 63.46\%  && 296,590 & 67.80\% && 153,708 & 35.14\% && 163,368 & 37.35\% \\ \hline \\\\

			\toprule
			\multicolumn{17}{c}{{\bf \sd}}\\\midrule
			 \multirow{2}{*}{\#Instances}  & \multicolumn{2}{c}{{n-gram} (c-match)} && \multicolumn{2}{c}{ {n-gram (p-match) }} 
			&& \multicolumn{2}{c}{ {CugLM (c-match) }} && \multicolumn{2}{c}{ {CugLM (p-match) }} 
			&&\multicolumn{2}{c}{ {T5 (non pre-trained) }} && \multicolumn{2}{c}{ {T5 (pre-trained) }} 
			\\ \cline{2-3} \cline{5-6} \cline{8-9} \cline{11-12} \cline{14-15} \cline{17-18}
			& {\#Correct} & {\%Correct} &&  {\#Correct} & {\%Correct}  && {\#Correct} & {\%Correct} && {\#Correct} & {\%Correct}  && {\#Correct} & {\%Correct} && {\#Correct} & {\%Correct}  \\ \hline
			457 & 20 & 4.37\% && 118 & 25.82\% && 214 & 48.35\% && 232 & 50.75\% && 142  & 31.07\% && 153 & 33.48\%  \\ \hline \\\\

			\toprule
			\multicolumn{17}{c}{{\bf \td}}\\\midrule
			\multirow{2}{*}{\#Instances}  & \multicolumn{2}{c}{{n-gram} (c-match)} && \multicolumn{2}{c}{ {n-gram (p-match) }} 
			&& \multicolumn{2}{c}{ {CugLM (c-match) }} && \multicolumn{2}{c}{ {CugLM (p-match) }} 
			&&\multicolumn{2}{c}{ {T5 (non pre-trained) }} && \multicolumn{2}{c}{ {T5 (pre-trained) }} 
			\\ \cline{2-3} \cline{5-6} \cline{8-9} \cline{11-12} \cline{14-15} \cline{17-18}
			& {\#Correct} & {\%Correct} &&  {\#Correct} & {\%Correct}  && {\#Correct} & {\%Correct} && {\#Correct} & {\%Correct}  && {\#Correct} & {\%Correct} && {\#Correct} & {\%Correct}  \\ \hline
			442 & 8 & 1.90\% && 66 & 14.93\% && 197 & 45.02\% && 209 & 47.28\% && 87 & 19.68\% && 101 & 22.85\% \\ \hline \\\\
			
		\end{tabular}
	\end{adjustbox}
\end{table*}
\end{landscape}

\begin{table*}[h]
	\centering
	\scriptsize
	\caption{McNemar's test (adj. $p$-value and OR) considering complete matches as correct predictions. P-t=pre-trained}
        \label{tab:stats-correct}
	\begin{tabular}{llrr}
		\toprule
		\textbf{Dataset} & \textbf{Test} & \textbf{\emph{p}-value} & \textbf{OR} \\ 
		\midrule
		\multirow{3}{*}{\fd} 
		& CugLM \emph{vs} T5 (p-t)  &$<0.001$  &3.54  \\ 
		& CugLM \emph{vs} $n$-gram &$<0.001$   &23.06  \\ 		
		& T5 (p-t) \emph{vs} $n$-gram &$<0.001$   &6.08   \\ \midrule
		
		\multirow{3}{*}{\sd}
		& CugLM \emph{vs} T5 (p-t)  &$<0.001$  &1.98  \\ 
		& CugLM \emph{vs} $n$-gram &$<0.001$   &98.0  \\ 		
		& T5 (p-t) \emph{vs} $n$-gram &$<0.001$   &10.50   \\\midrule		

		\multirow{3}{*}{\td}
		& CugLM \emph{vs} T5 (p-t)  &$<0.001$  &3.90  \\ 
		& CugLM \emph{vs} $n$-gram &$<0.001$   &32.50  \\ 		
		& T5 (p-t) \emph{vs} $n$-gram &$<0.001$   &24.25   \\		
		\bottomrule
	\end{tabular}
\end{table*}

\begin{figure*}[h!]
	\centering
	\includegraphics[width=\textwidth]{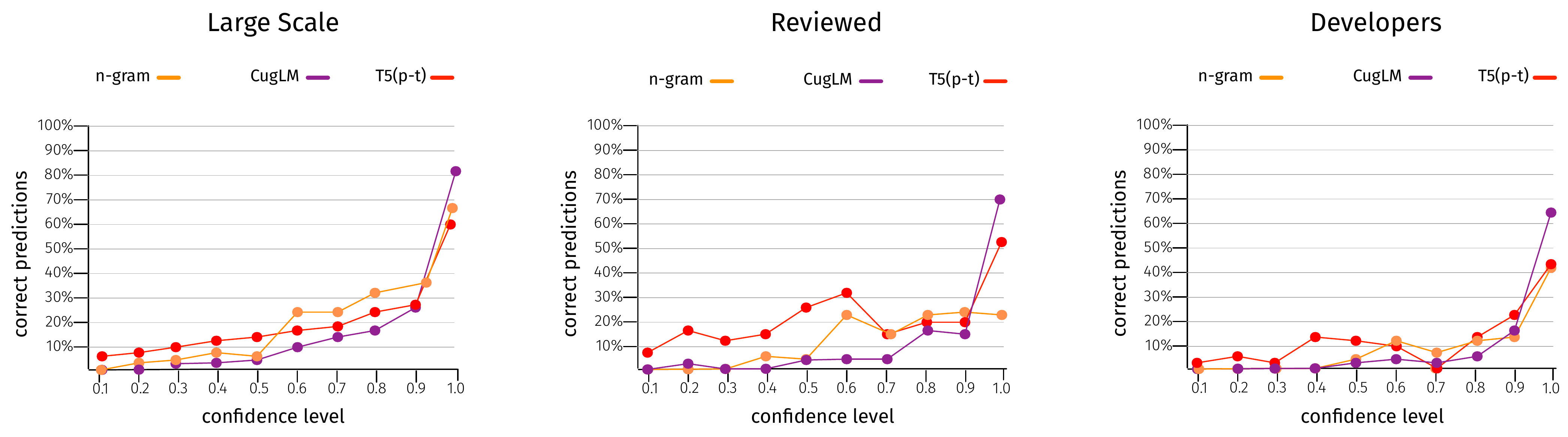}
	\caption{Percentage of correct predictions by confidence level}
	\label{fig:confidence}
\end{figure*}

\begin{figure*}[h!]
	\centering
	\includegraphics[width=\textwidth]{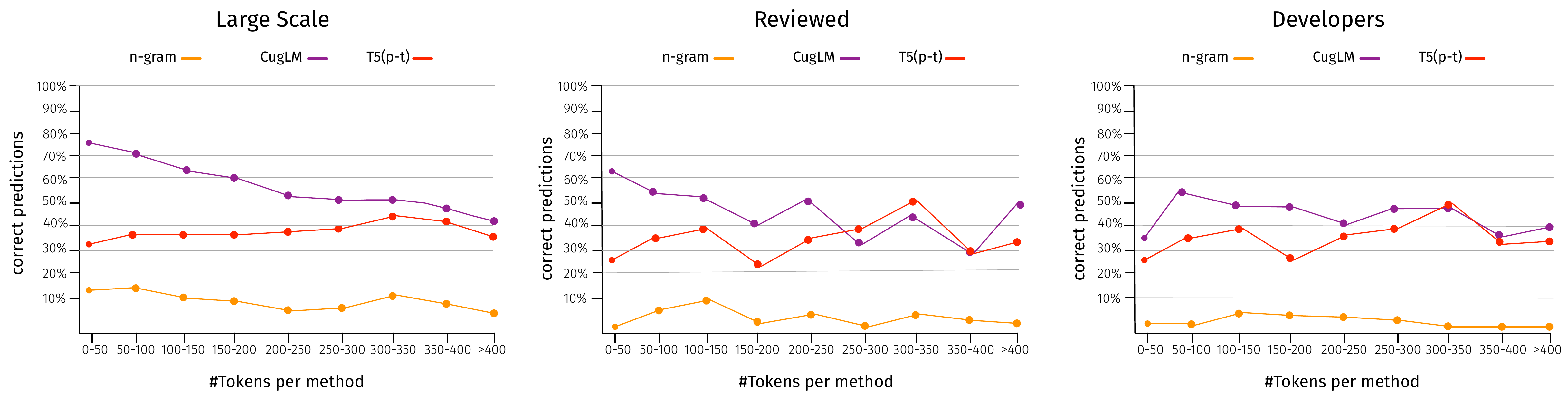}
	\caption{Percentage of correct predictions by tokens per method}
	\label{fig:prediction_length}
\end{figure*}

\figref{fig:confidence} depicts the relationship between the percentage of correct predictions and the confidence of the models. Similarly, to \figref{fig:prediction_length}, the orange line represents the $n$-gram model, while the purple and red lines represent CugLM and the T5 pre-trained model, respectively. Within each confidence interval (\eg 0.9-1.0) the line shows the percentage of correct predictions generated by the model (\eg $\sim$80\% of predictions having a confidence higher than 0.9 are correct for CugLM in the \fd). The achieved results show a clear trend for all models: Higher confidence corresponds to higher prediction quality. The best performing model (CugLM) is able, in the highest confidence scenario, to obtain {66\%} of correct predictions on the \td, {71\%} on the \sd, and {82\%} on the \fd. These results have a strong implication for the building of rename refactoring recommenders on top of these approaches: Giving the possibility to the user (\ie the developer) to only receive recommendations when the model is highly confident can discard most of the false positive recommendations.

Concerning the manual analysis we performed on 100 wrong recommendations generated by each model on the \fd, a few findings can be distilled. First, the three authors observed that the T5 was the one more frequently generating, in the set of wrong predictions we analyzed, identifiers that were meaningful in the context in which they were proposed (despite being different from the original identifier used by the developers). For example, \texttt{value} was recommended instead of \texttt{number} or \texttt{harvestTasks} instead of \texttt{tasks}. The three authors agreed on 31 meaningful identifiers proposed by the T5 in the set of 100 wrong predictions they inspected. Surprisingly, this was not the case for the other two models, despite the great performance we observed for CugLM. However, a second observation we made partially explains such a finding: We found that several failure cases of CugLM and of the $n$-gram model are due the recommendation of identifiers already used somewhere else in the method and, thus, representing wrong recommendations. We believe this is due to the different prediction mechanism adopted by the T5 as compared to the other two models. As previously explained, the T5 generates a single prediction for all instances of the identifier to predict, thus considering the whole method as a context for the prediction and inferring that identifiers already used in the context should not be recommended. 

The other two models, instead, scan the method token by token predicting each identifier instance in isolation. This means that if an identifier $x$ is used for the first time in the method after the first instance of the identifier $p$ to predict (\eg $p$ appears in line 2 while $x$ appears in line 7), the existence of $x$ is not considered when generating the prediction for $p$. This reduces the information available to CugLM and to the $n$-gram model. Also, CugLM has limitations inherited from the fixed size of its vocabulary set to the 50k most frequent tokens \cite{Liu:ase2020}, which are the only ones the model can predict. This means that CugLM is likely to fail when dealing with rare identifiers composed by several words. The T5, using the SentencePiece tokenizer, can instead compose complex identifiers.

\begin{figure*}[h!]
	\centering
	\includegraphics[width=\textwidth]{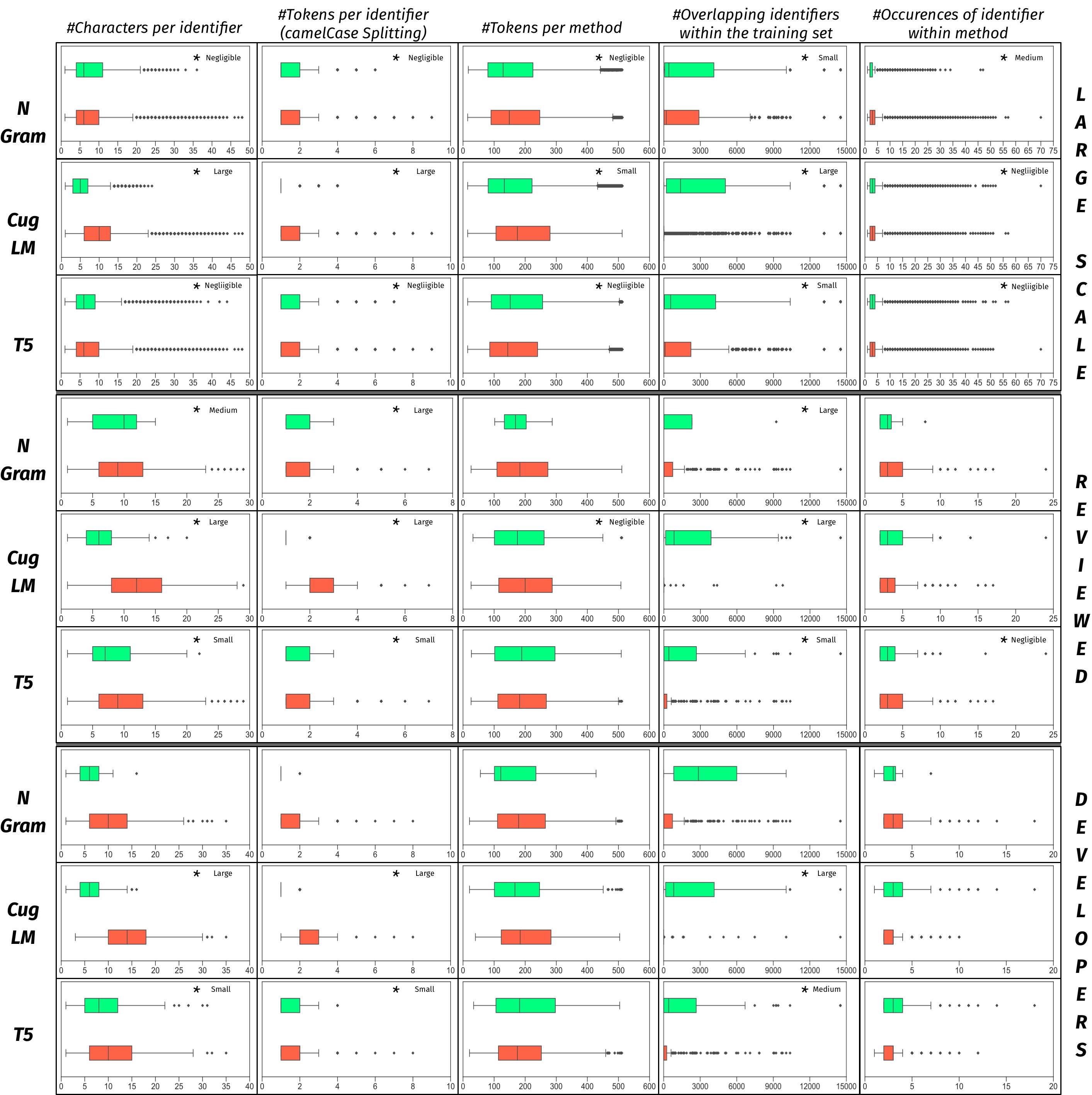}
	\caption{Characteristics of correct (green) and wrong (red) predictions}
	\label{fig:box}
\end{figure*}

Finally, \figref{fig:box} shows the comparison of five characteristics between correct (in green) and wrong (in red) predictions. The comparison has been performed on the three datasets (see labels on the right side of \figref{fig:box}), and for correct/wrong predictions generated by the three models (see labels on the left side of \figref{fig:box}). The characteristics we inspected are summarized at the top of \figref{fig:box}. 
For each comparison (\ie each pair of boxplots in \figref{fig:box}) we include a $\ast$ if the Mann-Whitney test reported a significant difference ($p$-value $<$ 0.05) and, if this is the case, the magnitude of the Cliff's Delta is reported as well.

Concerning the length of the target identifier (\emph{\#Characters per identifier} and \emph{\#Tokens per identifier}), the models tend to perform better on shorter identifiers, with this difference being particularly strong for CugLM. Indeed, this is the only model for which we observed a large effect size in the length of identifiers between correct and wrong predictions. Focusing for example on the length in terms of number of tokens (\emph{\#Tokens per identifiers}), it is clear that excluding rare exceptions, CugLM mostly succeeds for one-word identifiers. This is again likely to be a limitation dictated by the fixed size of the vocabulary (50k tokens) that cannot contain all possible combinations of words used in identifiers.

The size of the coding context (\ie method) containing the identifier to predict (\emph{\#Tokens per method}) does not seem to influence the correctness of the prediction, with few significant differences accompanied by a negligible or small effect size. \rev{
This is also visible by looking at \figref{fig:prediction_length}, which portrays the relationship between the percentage of correct predictions and the number of tokens composing the input method. The orange line represents the $n$-gram model, while the purple and red lines represent CugLM and the T5 pre-trained model, respectively. Within each interval (\eg 0-50), the line shows the percentage of correct predictions generated by the model for methods having a tokens length falling in that bucket. The only visible trend is that of \emph{CugLM} on the \fd, which shows a clear downward trend in correct predictions with the increase in length of the input method. This is indeed the only scenario for which the statistical tests reported a significant differences in the method length of correct and wrong predictions with a small effect size (in all other cases, the difference is not significant or accompanied by a small effect size).}

Differently, identifiers appearing in the training set tend to help the prediction (\emph{\#Overlapping identifiers within the training set}). This is particularly true for CugLM (large effect size on all datasets), since its vocabulary is built from the training set. The boxplot for the wrong predictions is basically composed only by outliers, with its third quartile equal 0. This indicates that the predictions on which CugLM fails are usually those for identifiers never seen in the training set. 

Finally, the number of times that an identifier to predict appears in the context (\ie \#Occurences of identifier within methods), only has an influence for the T5 on the \fd. However, there is no strong trend to discuss for this characteristic.

\subsection{Implications of our Findings} \label{sub:implications}
Our findings have implications for practitioners and researchers. For the first, our results show that modern DL-based techniques presented in the literature may be already suitable to be embedded in rename refactoring engines. Clearly, they still suffer of limitations that we will discuss later. However, especially when the confidence of their predictions is high, the generated identifiers are often meaningful, matching the ones chosen by developers.

In terms of research, there are a number of improvements these tools can benefit from. First, we noticed that the main weakness of the strongest approach we tested (\ie CugLM) is the fixed vocabulary size. Such a problem has been addressed in other models using tokenizers such as byte pair encoding \cite{bpe} or the SentencePiece tokenizer exploited by the T5. Integrating these tokenizers in CugLM (or similar techniques) could help in further improving performance. Second, we noticed that several false-positive recommendations could be avoided by just integrating into the models more contextual information. 

For example, if the model is employed to recommend an identifier in a given location $l$, other identifiers having $l$ in their scope do not represent a viable option, since they are already in use. Similarly, the integration of type information in CugLM demonstrated the boost of performance that can be obtained when the prediction model is provided with richer data. 	\rev{Also, the employed models are predicting identifier names without exploiting information such as (i) the original identifier name that could be improved via a rename refactoring, and (ii) the naming convention adopted in the project. Augmenting the context provided to the models with such information might substantially boost their prediction performance.}

Finally, while we performed an extensive study about the capabilities of data-driven techniques for variable renaming, our experiments have been performed in an ``artificial'' setting. The (mostly positive) achieved results encourage the natural next step represented by case studies with developers to assess their perceived usefulness of these techniques.
\section{Threats to Validity} \label{sec:threats}

\textbf{Construct validity.} Our study is largely based on one assumption: The identifier name chosen by developers is the correct one the models should predict. We addressed this threat when building two of our datasets: (i) we ensure that the variable identifiers in the \sd have been checked in the context of a code review process involving multiple developers; (ii) we built \td by looking for identifiers explicitly renamed by developers. Thus, it is more likely that those identifiers are actually meaningful. 

\textbf{Internal validity.} An important factor that influences DL performance is hyperparameters tuning. Concerning T5, for the pre-training phase we used the default T5 parameters selected in the original paper \cite{raffel2019exploring} since we expect little margin of improvement for such a task-agnostic phase. For the fine-tuning, due to feasibility reasons, we did not change the model architecture (\eg number of layers), but we experimented with different learning rates-scheduler as did before by Mastropaolo \etal~\cite{mastropaolo2021studying}. For the other two techniques we relied on the parameters proposed in the papers presenting them.

\textbf{External validity.} While the datasets used in our study represent hundreds of software projects, the main threat in terms of generalizability is represented by the focus on the \emph{Java} language. It is important to notice that the experimented models are language agnostic, but would require the implementation of different tokenizers to support specific languages. 

\section{Conclusion} \label{sec:conclusion}
We presented a large-scale empirical study aimed at assessing the performance of data-driven techniques for variable renaming. We experimented with three different techniques, namely the $n$-gram cached language model \cite{hellendoorn2017deep}, the T5 model \cite{raffel2019exploring}, and the Transformer-based model presented by Liu \etal \cite{Liu:ase2020}. We show that DL-based models, especially when considering predictions they generate with high confidence, represent a valuable support for variable rename refactoring. 
Our future research agenda is dictated by the implications discussed in \secref{sub:implications}.

\section{Data Availability} \label{sec:data-availability}

All the code and data we used in our study is publicly available in our replication package \cite{replication}. In particular, we provide: (i) the code needed to train the models, (ii) the datasets we built for training, validating, and testing the models, (iii) all predictions generated by the three different approaches, and (iv) our trained models.

\begin{acknowledgements}
This project has received funding from the European Research Council (ERC) under the European Union's Horizon 2020 research and innovation programme (grant agreement No. 851720).
\end{acknowledgements}

\newpage

\bibliographystyle{IEEEtran}
\bibliography{IEEEabrv,main}  

\end{document}